\documentclass[prd,showpacs,superscriptaddress,preprintnumbers,twocolumn,amsmath,nofootinbib]{revtex4}
 \usepackage[dvips,final]{graphicx}
  \usepackage{amssymb,amsmath,epsfig,bm,pifont}
   \graphicspath{{./figs/}}
\usepackage{color}

\usepackage{relsize}
\newcommand{\babar}{{\mbox{\slshape B\kern-0.1em{\smaller A}\kern-0.1em
            B\kern-0.1em{\smaller A\kern-0.2em R}}}
\def\MSbar{\relax\ifmmode\overline                                        
            {\rm MS}\else{$\overline{\rm MS}${ }}\fi}                     
           }
\begin{document}
\thispagestyle{empty}
 \date{\today}
  \preprint{\hbox{RUB-TPII-01/2015}}

\title{Chimera distribution amplitudes for the pion and the longitudinally polarized $\rho$-meson \\}

\author{N.~G.~Stefanis}
  \email{stefanis@tp2.ruhr-uni-bochum.de}
   \affiliation{Institut f\"{u}r Theoretische Physik II,
                Ruhr-Universit\"{a}t Bochum,
                D-44780 Bochum, Germany\\}

\author{A.~V.~Pimikov                            }
  \email{pimikov@theor.jinr.ru}
   \affiliation{Bogoliubov Laboratory of Theoretical Physics, JINR,
                141980 Dubna, Russia\\}
   \affiliation{Institute of Modern Physics, Chinese Academy of Sciences,
 	            Lanzhou, 730000, P. R. China\\}

\begin{abstract}
Using QCD sum rules with nonlocal condensates, we show that the
distribution amplitude of the longitudinally polarized $\rho$-meson
may have a shorttailed platykurtic profile in close analogy to our
recently proposed platykurtic distribution amplitude for the pion.
Such a chimera distribution de facto amalgamates the broad unimodal
profile of the distribution amplitude, obtained with a
Dyson--Schwinger equations-based computational scheme, with the
suppressed tails characterizing the bimodal distribution amplitudes
derived from QCD sum rules with nonlocal condensates.
We argue that pattern formation, emerging from the collective
synchronization of coupled oscillators, can provide a single
theoretical scaffolding to study unimodal and bimodal distribution
amplitudes of light mesons without recourse to particular computational
schemes and the reasons for them.

\end{abstract}

\maketitle

\section{Introduction}
\label{sec:intro}
Many theoretical models exist to describe the valence parton
distribution amplitude (DA) of the $\pi$, the $\rho$
(both longitudinally and transversally polarized),
and other mesons.
In particular the pion, the lightest bound state within Quantum
Chromodynamics (QCD), provides a suitable ``laboratory'' for testing
new ideas and techniques to catch the main ingredients of the
underlying quark-gluon dynamics entering exclusive QCD processes
(see \cite{Chernyak:1983ej,BL89,Stefanis:1999wy} for reviews).
In this paper we will present an amplification of the recent analysis
by one of us in \cite{Stefanis:2014nla},
continued in \cite{Stefanis:2014yha},
for the pion and consider its extension to the longitudinally
polarized $\rho$ meson ($\rho_\parallel$).
The new mode of thought in \cite{Stefanis:2014nla} is based on the
ubiquitous phenomenon of synchronization (Sync for short) in complex
systems and we will expand the status of this subject towards a deeper
understanding of the meson DAs.
In this way, we will redetermine the $\rho_\parallel$ DA using QCD
sum rules with nonlocal condensates (NLC) within the approach
developed in \cite{BMS01,BMS05lat,Pimikov:2013usa}.

Our primary findings to be discussed later can be summarized as
follows:
First, we provide more details about the structure of the
Sync-inspired shorttailed (i.e., endpoint-suppressed) platykurtic (pk)
$\pi$ DA, proposed in \cite{Stefanis:2014nla}, and quantify the
uncertainties of its expansion coefficients.
This DA is a kind of chimera\footnote{An imaginary creature in
Greek mythology made up of different animals.} state because
it mimics within the NLC-based approach characteristics
pertaining to the dynamical chiral symmetry breaking (DCSB)
described in terms of Dyson--Schwinger equations (DSE)
\cite{Chang:2013pq,Gao:2014bca}.
In particular it conserves NLC-generated endpoint suppression of the
$\pi$ DA combining it with a broad downward concave shape in the
central region.
Second, using similar mathematical techniques, we obtain within the
reliability range of the NLC approach a regime of DAs for the
$\rho_\parallel$ meson characterized by a shorttailed platykurtic
profile.
Third, we employ statistical measures, like the kurtosis, to classify
meson DAs with respect to their peakedness relative to the tail
flatness and heaviness.

The rest of the paper is organized as follows.
The next two sections (Sec.\ \ref{sec:DA-QCD} and
Sec.\ \ref{sec:gegen-repres}) discuss the theoretical basis for
the description of meson DAs within QCD.
Section \ref{sec:NLC-SR} sketches the derivation of the $\pi$ and
$\rho_\parallel$ DAs from QCD sum rules with NLCs.
We will then proceed to investigate how meson DAs can be analyzed in
terms of synchronization concepts (Sec.\ \ref{sec:sync}).
Synthetic meson DAs will be considered in Sec.\ \ref{sec:synthetic-DA},
where also the important chimera DAs with a shorttailed platykurtic
profile will be presented.
Finally, Sec.\ \ref{sec:concl} will be reserved for the summary of our
main results and conclusions.

\section{Meson DA\lowercase{s} in QCD}
\label{sec:DA-QCD}
Let us consider the pion DA, starting with its definition.
\footnote{The exposition to follow relies upon the review in
\cite{Stefanis:1999wy} to which we refer for details and the original
references.}
Applying collinear factorization in QCD, the $\pi$ DA of
leading-twist two,
$\varphi_{\pi}^{(2)}(x,\mu^2)$,
encodes the distribution of the longitudinal momentum of the pion
between its two valence constituents: the quark and the antiquark,
with corresponding longitudinal-momentum fractions
$x_q=x=(k^0+k^3)/(P^0+P^3)=k^+/P^+$
and
$x_{\bar{q}}=1-x\equiv \bar{x}$, respectively.
Its momentum-scale dependence stems from the renormalization of the
current operator in
\begin{eqnarray}
  \langle 0| \bar{q}(z) \gamma_\mu\gamma_5 [z,0] q(0)
           | \pi(P)
  \rangle|_{z^{2}=0}
&& \!\!\! \! \! =
  if_\pi P_\mu \int_{0}^{1} dx e^{i x (z\cdot P)}
\nonumber \\
&& \times \varphi_{\pi}^{(2)} \left(x,\mu^2\right) \, ,
\label{eq:pion-DA}
\end{eqnarray}
where the gauge link
$
 [z,0]
=
 \mathcal{P}\exp [
                  -ig \int_{0}^{z} dw_{\mu}A_{a}^{\mu}(w)t_{a}
                 ]
=
 1
$
is set equal to unity on account of the lightcone gauge
$A\cdot n\equiv A^+=0$ ($n^2=0$).
The pion DA is related to the Bethe--Salpeter wave function
$\psi_{\pi}(x,k_\perp)$
by integrating over the transverse parton momentum $k_\perp$, i.e.,
\begin{equation}
  \varphi_{\pi}^{(2)} \left(x,\mu^2\right)
\sim
  \int_{}^{k_{\perp}^2<\mu^2} d^2k_\perp \psi_{\pi}(x,k_\perp) \, .
\label{eq:wf}
\end{equation}
Because the dependence on the momentum scale of any meson DA is
controlled by the Efremov-Radyushkin--Brodsky-Lepage (ERBL)
\cite{Efremov:1978rn,BL80} evolution equation, each meson DA can be
expressed in terms of the one-loop eigenfunctions of this equation,
$\psi_n(x)=6x(1-x) C^{(3/2)}_n(2x-1)$,
with the asymptotic (asy) DA being given by
$\varphi_{\pi}^{\rm asy}(x)=6x(1-x)\equiv 6x\bar{x}$,
and $C^{(3/2)}_n(2x-1)$ denoting the Gegenbauer polynomials
of order $3/2$ within the complete and orthonormal basis on
$x\in[0,1]$ with respect to the weight $x\bar{x}$.
Thus, one has the (scale-dependent) conformal expansion
\begin{equation}
  \varphi_{\pi}^{(2)}(x,\mu^2)
=\sum_{n=0,2,4, \ldots}^{\infty} a_{n}(\mu^2) \psi_n(x)
\label{eq:gegen-exp}
\end{equation}
in terms of the nonperturbative coefficients $a_n(\mu^2)$.
By virtue of the normalization condition
$
 \int_{0}^{1}dx\varphi_{\pi}^{(2)}(x, \mu^2)
=
 1
$,
$a_0=1$ at any scale $\mu^2$.

\begin{figure}[t]
\centering
\includegraphics[width=0.40\textwidth]{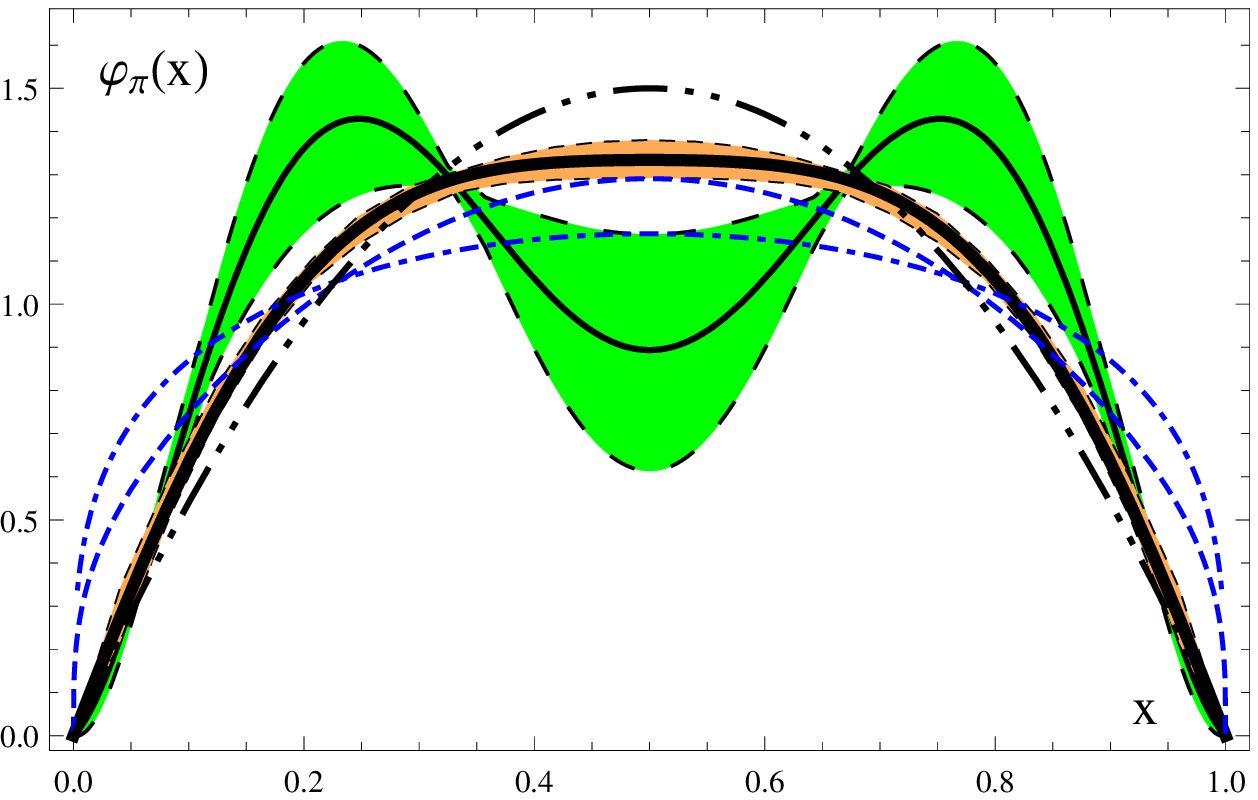} 
\hfill
\includegraphics[width=0.40\textwidth]{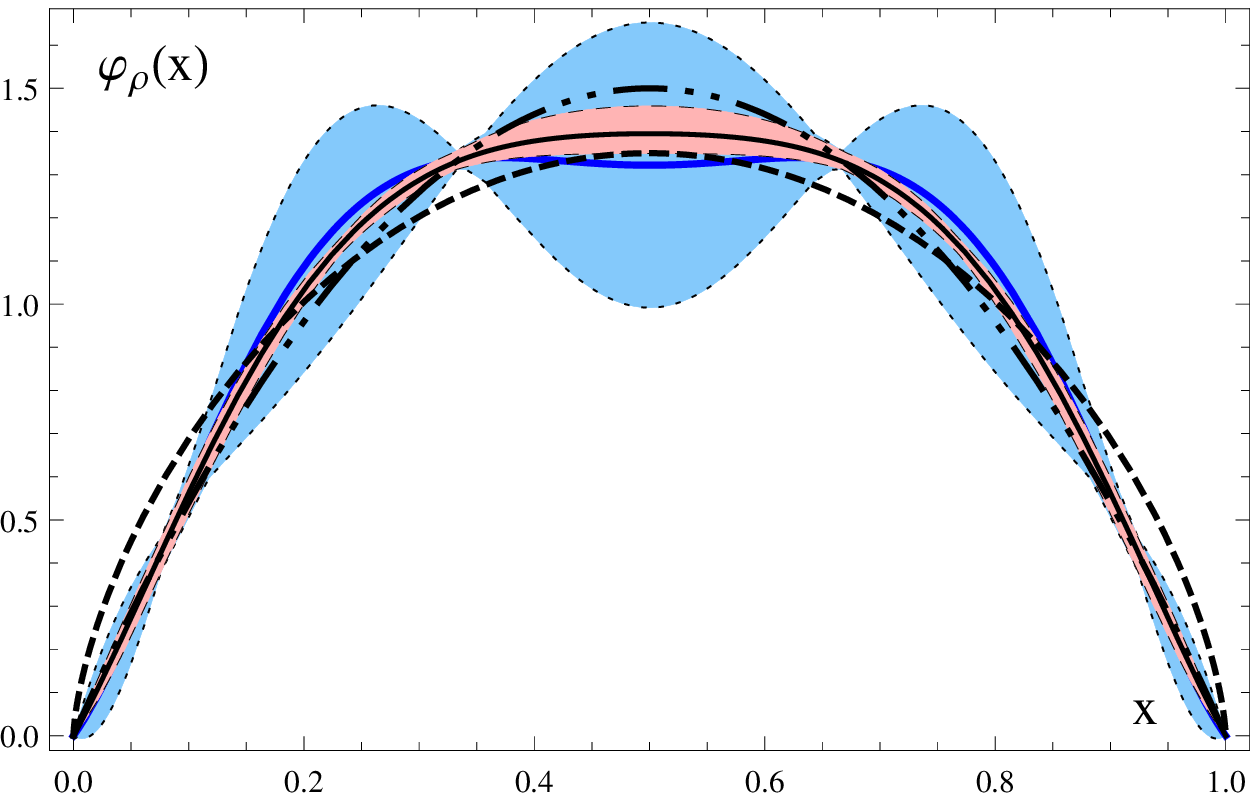} 
\caption{(color online). The upper panel shows some characteristic pion
DAs.
The larger shaded (green) area displays the region of the bimodal
Bakulev-Mikhailov-Stefanis (BMS) family of DAs derived in
\protect\cite{BMS01} from QCD sum rules with NLCs using the nonlocality
parameter
$\lambda_{q}^{2}=0.4$~GeV$^2$, defind in Sec.\ \ref{sec:gegen-repres}.
The family of the $\pi$ platykurtic DAs is shown as a narrow shaded
strip in red color, obtained with $\lambda_{q}^{2}=0.45$~GeV$^2$
at the edge of the NLC regime.
The solid lines within both regions denote, respectively, the BMS model
from \protect\cite{BMS01} and the pk DA discussed in the text and in
\protect\cite{Stefanis:2014nla,Stefanis:2014yha}.
The broken lines show the unimodal DSE-DB DA (dashed), the
DSE-RL DA (dashed-dotted) (both from \protect\cite{Chang:2013pq}),
and the asymptotic DA (dashed-dotted-dotted).
The lower panel illustrates various $\rho_{\parallel}$ DAs.
The larger blue shaded area contains the family of DAs obtained
in \protect\cite{Pimikov:2013usa} using QCD sum rules with NLCs and
$\lambda_{q}^{2}=0.4$~GeV$^2$, while the narrower strip in its interior
indicates the platykurtic regime of these DAs.
The solid lines within each band denote, respectively, the bimodal
DA from \protect\cite{Pimikov:2013usa} (lowest (blue) solid line)
and the platykurtic DA (upper solid line) derived in this work.
The lower dashed line represents the DA obtained from the DSE approach
\protect\cite{Gao:2014bca}, whereas the dashed-dotted-dotted line
displays again the asymptotic DA.
All DAs in both panels refer to the scale $\mu^2=4$~GeV$^2$ after
two-loop ERBL evolution, provided the initial
proprietary scale was lower than this.
\label{fig:pi-rho-DAs}}
\end{figure}

\section{Gegenbauer DA representations}
\label{sec:gegen-repres}
In our approach \cite{BMS01}, based on QCD sum rules with nonlocal
condensates \cite{MR86,MR89,BR91,MR90,MR92,BM98}, we calculated the
moments
\begin{equation}
  \langle \xi^{N} \rangle_{\pi}
\equiv
  \int_{0}^{1} dx (2x-1)^{N} \varphi_{\pi}^{(2)}(x,\mu^2)
\label{eq:moments}
\end{equation}
up to $N=10$ together with their intrinsic theoretical uncertainties at
the typical hadronic scale $\mu^2 \approx 1.35$~GeV$^2$ \cite{BMS01}.
Detailed estimates of the moment uncertainties can be found in
\cite{BMS04kg,Stefanis:2008zi} (see also Table \ref{tab:parameters})).
This scale represents the average value of the Borel parameter $M^2$ in
the stability window of the sum rule, notably,
$M^2 \sim [m_{\rho}^2, s_{0}]$,
where $s_0$ is the continuum threshold and $m_{\rho}$ is the physical
mass of the $\rho$ meson.
Note that the moments $\langle \xi^{N} \rangle_{\pi}$ coincide
by construction with the central moments
\begin{equation}
  \mu_{N}[\varphi_\pi]
=
  \int_{0}^{1} dx (x-\mu[\varphi_\pi])^N \varphi_\pi(x)
=
  2^{-N}\langle \xi^N\rangle_\pi
\label{eq:central-moments}
\end{equation}
of $\varphi_\pi(x)$, where
\begin{equation}
  \mu[\varphi_\pi]
=
  \int_{0}^{1} dx x \varphi_\pi(x)
=
  \frac{1}{2}
\label{eq:mean}
\end{equation}
is the mean of the DA.
Using standard techniques
(see, for example, \cite{Chernyak:1983ej,Stefanis:1999wy}),
we extracted from the moments the corresponding conformal coefficients
$a_n$ entering Eq.\ (\ref{eq:gegen-exp}).
These quantities contain nonperturbative information and implicitly depend
on the finite virtuality of the vacuum quarks, the latter expressed by means
of the nonlocality parameter
$
  \lambda_{q}^{2}
=
  \langle
  \bar{q}igG^{\mu\nu}\sigma_{\mu\nu}q \rangle / 2\langle
        \bar{q}(0)q(0)
  \rangle \approx [0.35-0.45]$~GeV$^2
$.
It was found that the first two coefficients
(see Appendix C in \cite{Stefanis:2012yw})
\begin{eqnarray}
\label{eq:conf-coeff-a2}
  a_2
& = &
  \frac{7}{12}\left(5\left\langle \xi^2 \right\rangle - 1\right) \, ,
\\
  a_4
& = &
  \frac{77}{8}\left(
                     \left\langle \xi^4 \right\rangle
                     -\frac{2}{3}\left\langle \xi^2 \right\rangle
                     +\frac{1}{21}
              \right)
\label{eq:conf-coeff-a4}
\end{eqnarray}
dominate.
Their values have been calculated with controlled accuracy in \cite{BMS01}.
They read $a_2(\mu^2\approx 1~{\rm GeV}^2)=0.20$ and
$a_4(\mu^2\approx 1~{\rm GeV}^2)=-0.14$,
whereas the next higher coefficients were computed in the same work as well
and found to be much smaller, viz.,
$a_6\approx a_2/3;
\quad a_8\approx a_2/4;
\quad a_{10}\approx a_2/5$,
but bearing large uncertainties.
Their inclusion can add refinements to the method, as we have
discussed in \cite{BMPS11}.
This apparent hierarchy, with each subsequent coefficient becoming
smaller with the order of the conformal expansion, is not
following from general principles; it is an inherent element of our
specific approach.
Indeed, one can even have an inverse ordering of the coefficients $a_n$
--- see \cite{ABOP10} for such DAs.
In the final analysis, the pion DA at the scale
$\mu^2 \gtrsim 1$~GeV$^2$
can be written in the form
($\xi\equiv 2x-1 = x-\bar{x}$)
\begin{equation}
  \varphi_{\pi}^\text{BMS}(x)
=
6x\bar{x}
 \left[
       1 + a_2 C_{2}^{(3/2)}(\xi) + a_4 C_{4}^{(3/2)}(\xi)
 \right] \, ,
\label{eq:truncated}
\end{equation}
where the label means Bakulev, Mikhailov, Stefanis \cite{BMS01}.
This simple model probably offers a biased picture of the pion
structure but its chief predictions are in good agreement with
measurements and various lattice simulations, as detailed in a series
of papers
\cite{BMS01,BMS02,Bakulev:2003cs,BMS05lat,BST07,Stefanis:2008zi,%
MS09,BMPS11,Bakulev:2012nh,Stefanis:2012yw,Mikhailov:2014rqa,Stefanis:2014yha}.
The reason is that physical observables, like the pion-photon
transition or the pion's electromagnetic form factor are given in
terms of integrals of the DAs with smooth coefficient functions.
Because the leading-order anomalous dimensions of the involved
operators in the matrix elements between the meson state and the
vacuum  are positive (except $\gamma_{0}=0$), the higher coefficients
are logarithmically suppressed at large scales, so that only
$\psi_{0}(x)=\varphi_{\pi}^{\rm asy}(x)$ survives.
This is particularly visible in the inverse moment of the pion DA,
cf. Eq. (\ref{eq:inv-mom}), in which the oscillating terms are
washed out and strongly suppressed as the momentum increases.
The profiles of the meson DAs, considered in this work, are shown
in the upper ($\pi$) and the lower ($\rho_{\parallel}$) panel of
Fig.\ \ref{fig:pi-rho-DAs}, respectively.

To leading logarithmic accuracy, the conformal coefficients $a_n$
are multiplicatively renormalizable and the anomalous dimensions
are known in closed form, see, for example, \cite{Stefanis:1999wy}.
At the next-to-leading-order (NLO) level, the momentum-scale
dependence of the conformal coefficients (or moments) is more
complicated owing to the mixing of the operators under renormalization
\cite{MR86ev,Mul94,Mul95}.
The diagonal elements of the corresponding two-loop anomalous-dimension
matrix coincide with the flavor nonsinglet anomalous dimensions known
from deeply inelastic scattering.
They have been computed in \cite{FRS77}, corrected in
\cite{GALY79}, and verified in \cite{Curci:1980uw}.
The NLO ERBL kernel was calculated in \cite{DR84,Sar84,MR85}, while the
analytic expressions for the mixing coefficients to obtain the
corresponding eigenfunctions were given in \cite{Mul94,Mul95}.
Bear in mind that the next-to-leading-order corrections under a change
of scale using a running coupling appear as a two-loop contribution
of the eigenvalues and as an $\alpha_s$ correction to the
eigenfunctions.
A detailed exposition of the ERBL evolution of the pion DA at the
two-loop level, as used in the present work, is provided in Appendix
D in \cite{BMS02}.

It is convenient to employ another Gegenbauer representation
(``Gegenbauer-$\alpha$'')
proposed in \cite{Chang:2013pq,Gao:2014bca}, notably,
\begin{eqnarray}
&&  \varphi_{\pi}^{(2)}(x, \mu^2)
=
  f(\{\alpha,a^\alpha_2,...,a^\alpha_{j_s}\},x)
=
  \psi^{(\alpha)}_0(x)
\nonumber \\
&&  \quad + \sum_{j=2,4,\ldots}^{j_{s}}
        a_{j}^{\alpha}(\mu^2)\psi^{(\alpha)}_n(x) \, ,
\label{eq:DA-gen-Gegen}
\end{eqnarray}
where
\begin{equation}
  \psi^{(\alpha)}_n(x)
=
 N_{\alpha}(x\bar{x})^{\alpha_{-}}C^{(\alpha)}_{n}(2x-1)
\label{eq:gen-exp}
\end{equation}
and
$N_{\alpha}=1/B(\alpha + 1/2, \alpha + 1/2)$,
$\alpha_{-}=\alpha -1/2$, where $B(x,y)$ is the Euler beta function.
The Gegenbauer polynomials $C^{(\alpha)}_{n}(2x-1)$ form an
orthonormal set over $x\in[0,1]$ with respect to the weight
$[x\bar{x}]^{\alpha_{-}}$.
The difference to the conformal expansion in Eq.\ (\ref{eq:gegen-exp})
is that the order of the Gegenbauer polynomials is not a priori fixed
to the value $3/2$, but is allowed to vary in order to accelerate
the reconstruction procedure of meson DAs on $x\in[0,1]$.
However, expansion (10) is not directly amenable to ERBL evolution because the
functions $\psi_{n}^{\alpha}(x)$ are not its eigenfunctions.
To evolve $\varphi_{\pi}^{(2)}(x, \mu^2)$, expressed via
(\ref{eq:DA-gen-Gegen}), to another scale $Q^2>\mu^2$, one has to
project it first onto the conformal basis $\{\psi_{n}(x)\}$ and then
determine $\alpha_{-}$ and $a_{j}^{\alpha}$ at the new scale.
The authors of the works in Refs. \cite{Chang:2013pq,Gao:2014bca}
find that it is sufficient to include only one coefficient in this
expansion, namely, $a_{2}^{\alpha}$, so that
Eq.\ (\ref{eq:DA-gen-Gegen}) reduces to
\begin{equation}
  \varphi_{\pi}^{(\alpha)}(x,\mu^2)
=
  N_\alpha (x\bar{x})^{\alpha_{-}}
  [1 + a_{2}^{\alpha}C_{2}^{(\alpha)}(x-\bar{x})] \, .
\label{eq:DSE-DA}
\end{equation}
Below, we will use for our analysis both representations in parallel
and present the results in the form
$(a_2, a_4)$ and $\left(N_{\alpha}, \alpha_{-}, a_{2}^{\alpha}\right)$.
It is worth bearing in mind that broad DAs of the form
$\varphi_{\pi}(x) \sim (x\bar{x})^{\alpha_{-}}$
are not well represented by Eq.\ (\ref{eq:truncated}) which only
employs the first two conformal coefficients.
To approximate such DAs with sufficient accuracy, one would have to
include a large number of coefficients of order 50 or more,
see, for example, \cite{Mul94}.

\begin{figure*}[t]
\centering
\includegraphics[width=0.45\textwidth]{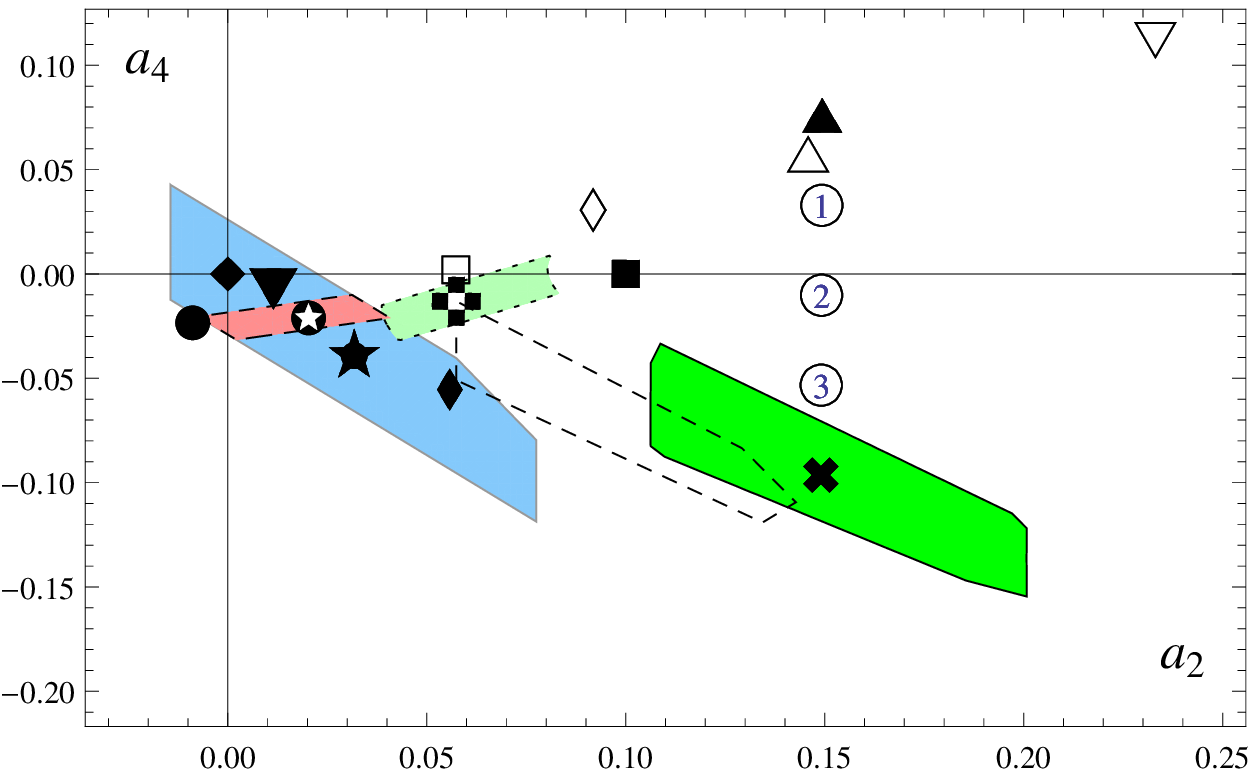} 
\hfill
\includegraphics[width=0.436\textwidth]{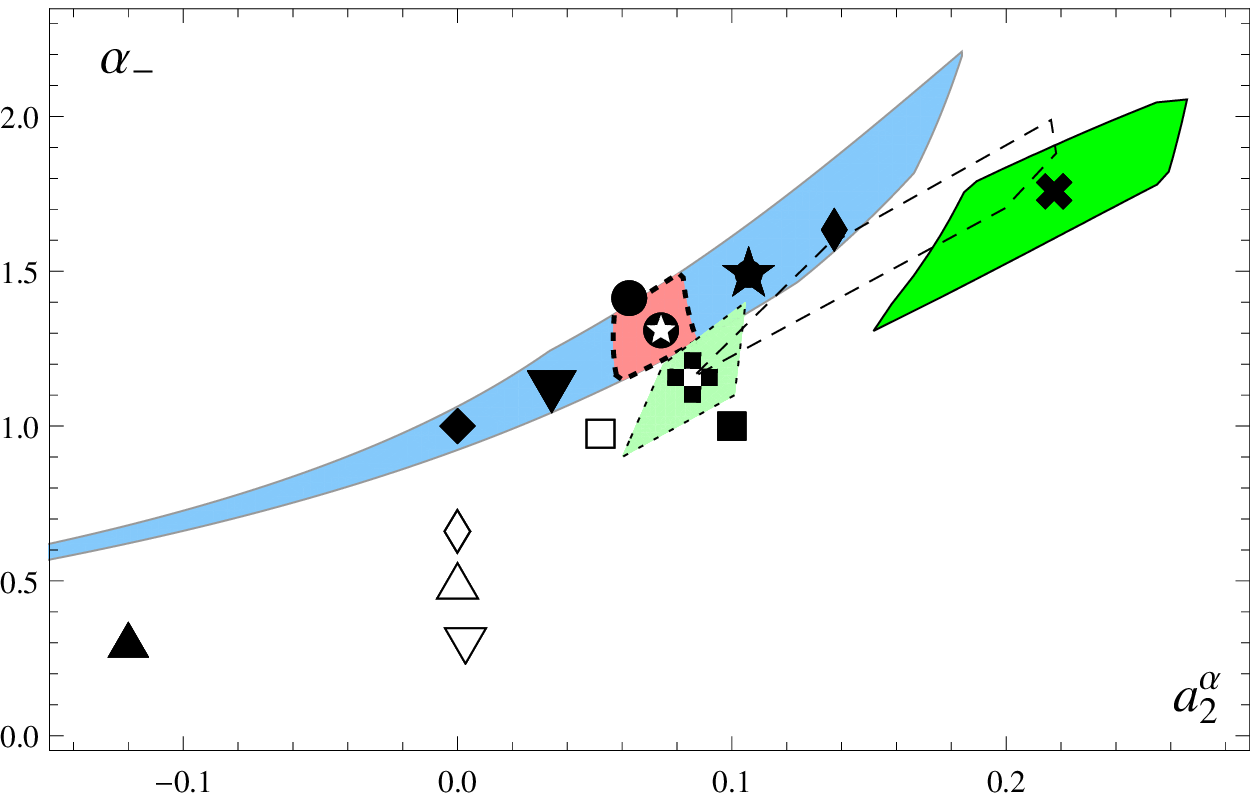} 
\caption{
(color online). Computed locations of various $\rho_{\parallel}$ and
$\pi$ DAs at $\mu^2=4$~GeV$^2$ in the plane spanned by the conformal
coefficients $a_2$ and $a_4$ (left panel) and corresponding results
in the plane $(\alpha_{-}, a_{2}^{\alpha})$ (right panel).
The slanted (blue) rectangle further to the left shows the region of
$\{a_2, a_4\}$ values obtained with NLC-QCD sum rules and
$\lambda_{q}^{2}=0.4$~GeV$^2$ for the $\rho_{\parallel}$ DA in
\protect\cite{Pimikov:2013usa}, with the smaller (red) rectangle in its
interior denoting the platykurtic regime determined in this work.
The other small (light-green) rectangle next to it contains the
platykurtic region calculated for the pion DA in this work using
$\lambda_{q}^2=0.45$~GeV$^2$ at the edge of the NLC regime.
The larger slanted (green) rectangle further to the right contains
the original values of $\{a_2,a_4\}$ for the pion DA obtained with
$\lambda_{q}^2=0.4$~GeV$^2$ in \protect\cite{BMS01}.
The symbols \ding{172}, \ding{173}, and \ding{174} show
examples of synthetic DAs according to Eq.\ (\ref{eq:hybrid}) for
$a=0.25$, $a=0.5$, and $a=0.75$, respectively.
Corresponding results for the $\rho_\parallel$ DA at 4~GeV$^2$ in
the $(\alpha_{-},a_2^{\alpha})$ plane are displayed in the right panel
with the following designations:
The stretched (blue) strip shows the region of
$(\alpha_{-}, a_{2}^{\alpha})$
values calculated with NLC-QCD sum rules and
$\lambda_{q}^2=0.4$~GeV$^2$ in this work.
The two smaller slanted rectangles illustrate the platykurtic regime
for the $\rho_{\parallel}$ DA (upper red rectangle) and for the $\pi$
DA (lower light-green rectangle); they correspond to the analogous
graphs in the left panel.
The other symbols in both panels denote the following DAs:
\ding{72} --- bimodal $\rho_{\parallel}$ DA from NLC-QCD sum rules
\protect\cite{Pimikov:2013usa};
$\blacklozenge$ --- bimodal $\rho_{\parallel}$ DA from NLC-QCD
sum rules \protect\cite{BM98};
\ding{74} --- platykurtic $\rho_{\parallel}$ DA (this work);
\ding{116} --- $\rho_{\parallel}$ DA from lightfront model
\protect\cite{Choi:2007yu}:
\ding{108} --- $\rho_{\parallel}$ DA from instanton model
\protect\cite{Dorokhov:2006xw};
\ding{110} --- $\rho_{\parallel}$ DA from QCD sum rules
\protect\cite{Ball:2007rt};
$\bm{\square}$ --- $\rho_{\parallel}$ DA from AdS/QCD
\protect\cite{Ahmady:2012dy}
$\bigtriangleup$ --- $\pi$ AdS/QCD \protect\cite{BT07};
$\bm{\lozenge}$ --- $\rho_{\parallel}$ DSE DA \protect\cite{Gao:2014bca};
\ding{117} --- asymptotic DA;
\ding{54} (both panels) --- BMS $\pi$ DA \protect\cite{BMS01};
\ding{60} (both panels) --- platykurtic $\pi$ DA;
\protect\cite{Stefanis:2014nla};
\ding{115} (both panels) --- DSE-DB $\pi$ DA
\protect\cite{Chang:2013pq};
$\bigtriangledown$ (both panels) --- DSE-RL $\pi$ DA
\protect\cite{Chang:2013pq}.
The slanted rectangles, bounded by dashed lines,
in both panels display the regions
determined in this work for the $\pi$ DA using NLC QCD sum rules with
$\lambda_q^2=0.45$~GeV$^2$.
When the proprietary scale of the studied DAs was different from
2~GeV, NLO evolution was applied to determine their coefficients at
this scale.
The numerical values of all coefficients are given in
Table \ref{tab:parameters}.
\label{fig:rhoL}}
\end{figure*}

\section{$\pi$ and $\rho_\parallel$ DA\lowercase{s} from NLC QCD sum rules}
\label{sec:NLC-SR}
In this section, we discuss the derivation of the
$\rho_{\parallel}$ DA using QCD sum rules with nonlocal condensates
within the scheme developed in \cite{BMS01,BMS05lat,Pimikov:2013usa}.
The main conceptual idea is to apply the sum-rule method in
combination with vacuum averages of nonlocal operators
\cite{MR86,MR89,BR91,MR90,MR92,BM98}.
Following this rationale, one can determine the moments
$\langle \xi^{N} \rangle$ of a meson DA using a sum rule with
nonlocal condensates.
In addition, due to the absence of endpoint singularities in the NLC
approach, one is able to calculate the inverse moment
\begin{equation}
  \langle
         x^{-1}
  \rangle_\pi
\equiv
  \int_{0}^{1}dx x^{-1}\varphi_{\pi}(x)
\label{eq:inv-mom}
\end{equation}
via an independent sum rule at the same low renormalization scale
$\mu \gtrsim 1$~GeV.
This moment is an integral characteristic of the pion DA and
encodes information on the maximum possible weight of the
higher-order conformal coefficients.
Moreover, it is particularly relevant for phenomenological applications
because
\begin{equation}
  \langle x^{-1}\rangle_{\pi}
=
  3(1+a_{2}+a_{4} \ldots)
=
  \frac{3}{\sqrt{2}f_{\pi}}Q^2F_{\gamma^*\gamma\pi^0}^{(\rm LO)}(Q^2)\, ,
\label{eq:diagonal}
\end{equation}
where $F_{\gamma^*\gamma\pi^0}^{(\rm LO)}(Q^2)$
is the leading-order (LO) expression of the pion-photon transition form
factor, which has been measured in several experiments from a few
GeV$^2$ up to 40~GeV$^2$ \cite{CELLO91,CLEO98,BaBar09,Belle12}.
Here the ellipsis represents corrections due to higher eigenfunctions
and evolution of the Gegenbauer coefficients $a_n$ to the considered
scale $Q^2$ is assumed (see \cite{BMS01} for more details).
Thus, experimental evidence can be used to validate or reject
particular pion DAs.
For instance, we know (see, e.g., \cite{Bakulev:2012nh}) that
all existing data demand $\langle x^{-1} \rangle_\pi >3$, implying that
the $\pi$ DA has to be broader than $\varphi_{\rm asy}$
at accessible momentum values.

To obtain the $\pi$ DA, we employ the following QCD sum rule
($m_\pi=0$, $f_{\pi}=0.137$~GeV)
\begin{eqnarray}
&& \!\!\!\!\!f_{\pi}^2\,\varphi_\pi(x) e^{-m_{\pi}^2/M^2}
  + f_{A_{1}}^2\,\varphi_{A_{1}}\!(x)\, e^{-m_{A_{1}}^2/M^2}
\nonumber \\
&& \quad =
  \int\limits_{\!\!0}^{~~s_{0}^{A_{1}}}
  \!\!\rho_\text{pert}\left(s,x\right)
  e^{-s/M^2}\!\!ds
  + \Phi_{\pi}(x,M^2) \, ,
\label{eq:SRpiconden}
\end{eqnarray}
whereas for the $\rho_{\parallel}$ DA we have
\begin{eqnarray}
&& \!\!\!\!\!f_{\rho}^2\,\varphi^{\parallel}_\rho(x)e^{-m_{\rho}^2/M^2}
  + f_{\rho'}^2\,\varphi^{\parallel}_{\rho'}\!(x)\, e^{-m_{\rho'}^2/M^2}
\nonumber \\
&& \quad =
  \int\limits_{\!\!0}^{~~s_0}
  \!\!\rho_\text{pert}\left(s,x\right)
  e^{-s/M^2}\!\!ds
  + \Phi_\rho(x,M^2)
\label{eq:SRrhoLconden}
\end{eqnarray}
with
$m_{\rho}=0.775$~GeV and $m_{\rho^{\prime}}=1.496$~GeV.
The effective $A_1$-meson state with the decay constant
$f_{A_{1}}=0.227$~GeV and the mass $m_{A_{1}}=1.2712$~GeV
comprises the $\pi^\prime$ and the $a_1$ mesons and is described by
the DA $\varphi_{A_{1}}(x)$.
The perturbative contribution to the sum rules is expressed
via the perturbative spectral density for which we use the corrected
expression published as Eq.\ (1) in the Erratum to \cite{BMS01}, viz.,
\begin{equation}
  \rho_{\rm pert}^{\rm (NLO)}(x)
=
  \frac{3x\bar{x}}{2\pi^2} \left[
                  1 + \frac{\alpha_s(\mu^2)}{4\pi}C_{\rm F}
                  \left(
                        5-\frac{\pi^2}{3} + \ln^2\frac{\bar{x}}{x}
                  \right)
            \right] \, ,
\label{eq:spec-dens-NLO}
\end{equation}
where $C_{\rm F}=(N_c^2-1)/2N_c=4/3$ for
$SU(3)_c$.\footnote{This expression coincides with the
$\mathcal{O}(\alpha_s)$ radiative correction entering the
perturbative contribution to the sum rule considered in
\cite{MR92}.}
The nonperturbative content of the sum rule is contained in the
expression
\begin{eqnarray}
&& \Phi_{\rho(\pi)}(x,M^2)
=
  \mp \Phi_\text{4Q}(x,M^2)
  + \Phi_{\bar q Aq}(x,M^2)
\nonumber \\
&& \quad \quad + \Phi_\text{V}(x,M^2)\! + \Phi_\text{G}(x,M^2) \, ,
\label{eq:SRrhoLterms}
\end{eqnarray}
where $M^2$ denotes the Borel parameter and $s_0$ marks in each case
the threshold value which separates the lowest resonance state from
higher states.
To saturate the sum rules for the first $N=10$ moments of
$\varphi_{\rho(\pi)}(x)$
\cite{BMS01}, we use $s_{0} \approx 2.25$~GeV$^2$.
The various contributions, forming $\Phi_{\rho(\pi)}$, pertain to
the following terms:
(i) $\Phi_{\text{4Q}}$ (four-quark condensate);
(ii) $\Phi_{\bar q Aq}$ (quark-gluon-antiquark condensate)
(iii) $\Phi_\text{V}$ (vector quark condensate);
(iv) $\Phi_\text{G}$ (gluon condensate).
Their explicit expressions can be found in Appendix A of
Ref.\ \cite{MPS10}.
The basic assumption here is that higher-order correlations are less
important than two-particle correlations
(vacuum-dominance hypothesis \cite{SVZ1}).
This assumption is employed in order to reduce the four-quark
condensate to the product of two-quark condensates ignoring
corresponding uncertainties.
One notes that the four-quark contribution enters the sum rule for
$\rho_{\parallel}$ in (\ref{eq:SRrhoLterms}) with the opposite sign
with respect to $\pi$.
As a result, it reduces the relative weight of this condensate in
the sum rule entailing smaller values of the DA moments --- in contrast
to the pion case.
In fact, we found in \cite{Pimikov:2013usa} (see Table 1 there) the
following relation:
$
 \langle \xi^{2N} \rangle_{\pi}
\geq
 \langle \xi^{2N} \rangle_{\rho_{\parallel}}
\geq
 \langle \xi^{2N} \rangle_{\rm asy}
$
with $N=1,2,3, \ldots$.

Similarly to the pion DA, the DA of the longitudinally polarized $\rho$
is defined by the matrix element
\begin{eqnarray}
  \langle 0|\bar{d}(z) \gamma_\mu u(0)| \rho(p,\lambda) \big|_{z^2=0}
&& \!\!\! \! \! =
  f_{\rho}^{\parallel} p_\mu\,
  \int_{0}^{1} dx e^{ix (z \cdot p)}
\nonumber \\
\mathrm{\mathrm{}}&& \times
  \varphi_\rho^{\parallel}(x,\mu^2) \, ,
\label{def:rhoL}
\end{eqnarray}
while the definition of the transversal ($\perp$) $\rho$ DA reads
\begin{eqnarray}
&&  \langle 0| \bar d(z) \sigma_{\mu\nu}u(0)| \rho(p,\lambda)\big|_{z^2=0}
  =
  i f_\rho^{\perp}(\varepsilon^{(\lambda)}_\mu p_\nu-\varepsilon^{(\lambda)}_\nu p_\mu)
\nonumber \\
&& \quad \times \int_{0}^{1} dx e^{ix (z \cdot p)}
 \varphi_\rho^{\perp}(x,\mu^2) \, ,
\label{def:rhoT}
\end{eqnarray}
where we have again employed the gauge $A^+=0$.
The $\rho_{\perp}$ DA will not be considered in this work.

\begin{table*}
\begin{center}
\begin{ruledtabular}
\caption{Various parameters entering the DA Gegenbauer representations
in (\ref{eq:gegen-exp}) and (\ref{eq:DA-gen-Gegen}) in the
$\overline{\rm MS}$ scheme.
The kurtosis $\beta_2$, the second moment
$\langle \xi^2\rangle$, the fourth moment $\langle \xi^4 \rangle$, and
the inverse moment $\langle x^{-1}\rangle$ of the $\pi$ and the
$\rho_{\parallel}$ DAs are also shown.
Always the original functional forms of the DAs have been used
with the same designations as in Fig.\ \ref{fig:rhoL}.
The reference scale for all entries is $\mu^2=4$~GeV$^2$ either by
construction or after NLO evolution.
Only the central values of the normalization coefficients $N_\alpha$
are displayed.
Note that we ignore here and in the figures the numerically
negligible effects on the asymptotic DA induced by NLO evolution
\cite{Mul95}.
\label{tab:parameters}}
\smallskip
\smallskip
\begin{tabular}{llllllllll}
\noalign{\smallskip}
Model DA                                   &$a_2$    &$a_4$     &$N_\alpha$     &$\alpha_{-}$   &$a_{2}^{\alpha}$ &$\beta_{2}$
                                                                                                                  & $\langle \xi^2\rangle$
                                                                                                                  & $\langle \xi^4\rangle$
                                                                                                                  & $\langle x^{-1}\rangle$
                                                                                                                                    \\[3pt]
\noalign{\smallskip}\hline
    asy  \ding{117}                        & $0$   & $0$   & $6$   & $1$   & $0$   & $2.14$
																						    & $0.2$ & $0.086$
																						    & $3$     \\
$\rho_{\parallel}$~\cite{BM98} $\blacklozenge$
                                           & $0.056$ & $-0.055$ & $16.83$       & $1.63$        & $0.137$  & $1.93$  & 0.22 & 0.093   & $3.0$   
                        \\
$\rho_{\parallel}$ \cite{Pimikov:2013usa} \ding{72} 
						& 0.032(46)
						& -0.038(81)
						& $13.6$
						& $1.50_{-1.50}^{+0.71}$
						& $0.11_{-1.14}^{+0.08}$
						& $2.0(3)$
						& 0.211(16)
						& 0.088(7)
						& $3.0(1)$
						\\
$\rho_{\parallel}^{\rm pk}$ (here) \ding{74}  
& $0.017(24)$                 
& $-0.021(11)$                
& $10.0$
& $1.312_{-0.171}^{+0.19}$    
& $0.071_{-0.015}^{+0.016}$   
& $2.06(4)$                   
& 0.206(8)                    
& 0.087(6)                    
& $3.0(8)$                    
\\
$\rho_{\parallel}$~\cite{Dorokhov:2006xw} \ding{108}    & $-0.009$ & $-0.023$ & $11.81$ & $1.41$  & $0.063$ & $2.09$ & 0.197 & 0.081 & $2.92$    \\
$\rho_{\parallel}$~\cite{Choi:2007yu} \ding{116}        & $0.012$  & $-0.007$ & $7.18$  & $1.11$  & $0.034$ & $2.11$ & 0.204 & 0.088 & $2.98$    \\
$\rho_{\parallel}$~\cite{Ball:2007rt} \ding{110}        & $0.10$   & $0$      & $6$     & $1$     & $0.10$  & $1.98$ & 0.234 & 0.109 & $3.30$    \\
$\rho_{\parallel}$~\cite{Gao:2014bca} $\bm{\lozenge}$   & $0.092$  & $0.031$  & $3.37$  & $0.66$  & $0$     & $2.05$ & 0.232 & 0.110 & $3.5$     \\
$\rho_{\parallel}$~\cite{Ahmady:2012dy} $\square$       & $0.057$  & $0.002$  & $5.76$  & $0.975$ & $0.052$ & $2.05$ & 0.220 & 0.099 & $3.2$     \\
$\pi_{\rm BMS}$~\cite{BMS01} \ding{54}
						& $0.149_{-0.043}^{+0.052} $
						& $-0.096_{-0.058}^{+0.063}$
						& 20.49
						& $1.76_{-0.45}^{+0.30} $
						& $0.217_{-0.066}^{+0.048}$
						& $1.74_{-0.14}^{+0.16} $
                        & $0.248_{-0.015}^{+0.016}$
                        & $0.108_{-0.03}^{+0.05}$
						& $3.16_{-0.09}^{+0.09}$
                        \\
$\pi_{\rm pk}$~\cite{Stefanis:2014yha} \ding{60}
                        & $0.057_{-0.019}^{+0.024}$   
						& $-0.013_{-0.019}^{+0.022}$  
						& 7.78                        
						& $1.16_{-0.26}^{+0.24}$      
						& $0.086_{-0.026}^{+0.019}$   
						& $2.02_{-0.03}^{+0.02}$      
                        & $0.220_{-0.006}^{+0.009}$   
                        & $0.098_{-0.005}^{+0.008}$   
						& $3.13_{-0.10}^{+0.14}$      
						\\
$\pi_{\rm DSE-DB}$~\cite{Chang:2013pq} \ding{115}         & $0.149$           & $0.076$ & $1.81$  & $0.31$  & $-0.12$  & $2.0$  & 0.251 & 0.128  & $4.6$   \\
$\pi_{\rm DSE-RL}$~\cite{Chang:2013pq} $\bigtriangledown$ & $0.233$           & $0.112$ & $1.74$  & $0.29$  & $0.0029$ & $1.9$  & 0.280 & 0.151 & $5.5$   \\
$\pi_{\rm AdS/QCD}$~\cite{BT07} $\bigtriangleup$          & $0.107$           & $0.038$ & $2.55$  & $0.50$  & $0$      & $2.03$ & 0.237 & 0.114 & $4.0$
\footnote{This value was obtained using only the first three terms of the conformal expansion in Eq.\ (\ref{eq:gegen-exp})
and is therefore not a precise estimate.}
   \\
$\pi_{\rm CZ}$~\cite{Chernyak:1983ej}                     & $0.42$              & $0$  & $6.0$  & $1.0$   & $0.42$   & $1.54$ & 0.343    & 0.181 & $4.25$  \\
$\pi_{\rm lat}$~\cite{Braun:2015axa}                      & $0.1364(154)(145)$  & --  & --  &  --  & --   & --  & $0.2361(41)(39)$~ & -- & --      \\
\noalign{\smallskip}
\end{tabular}
\end{ruledtabular}
\end{center}
\end{table*}

To construct $\varphi_\rho^{\parallel}(x,\mu^2)$, we compute the
moments $\langle \xi^N \rangle_{\rho_{\parallel}}$ up to $N=10$ from
the QCD sum rule (\ref{eq:SRrhoLconden}) and determine from them
the corresponding conformal coefficients
$a_n$ with $N=0,2,\ldots,10$.
Because the moments with $N\geq6$ turn out to have values close to the
asymptotic ones, we can safely set the associated conformal
coefficients equal to zero and express
$\varphi_\rho^{\parallel}(x,\mu^2)$
in the form of Eq.\ (\ref{eq:truncated}).
In addition, we use the Gegenbauer-$\alpha$ expansion, given by
Eq.\ (\ref{eq:DA-gen-Gegen}), via the parameters
$(\alpha_{-}, a_{2}^{\alpha})$.
The accessible regions of these parameters for the longitudinal $\rho$
DA, determined within QCD sum rules with nonlocal condensates, are
displayed as shaded blue areas closer to the $y$ axis in
Fig.\ \ref{fig:rhoL}.
The left panel displays the results in the $(a_2, a_4)$ plane, while
the right panel contains the analogous results in the
$(\alpha_{-}, a_{2}^{\alpha})$ plane.
We have also depicted in both panels the parameter regions referring
to the pion case for $\lambda_{q}^2=0.4$~GeV$^2$ (larger green
slanted rectangles further to the right) and for
$\lambda_{q}^2=0.45$~GeV$^2$
(transparent rectangles within dashed boundaries).
The graphics in Fig.\ \ref{fig:rhoL} include the platykurtic regimes of
both DAs, determined in our present NLC-based analysis to be outlined
later.
Those areas referring to $\rho_{\parallel}$ have red color and are
located closer to the ordinate (left panel) and further to the top
(right panel).
The analogous areas for the pion appear in light green color and are
adjacent to the previous ones.
In addition, we incorporate several other $\pi$ and $\rho_{\parallel}$
DAs with individual designations explained in the figure caption for
the readers's convenience.
The uncertainties of the presented pion and rho-meson DAs, obtained with
our NLC QCD SR approach, include only those stemming from the SRs
themselves and are related to the variation of the Borel parameter
within the stability window of the SRs \cite{BMS01}.
Experimental uncertainties in the input physical parameters have little
influence on the results and have been ignored.

The values of all parameters of the displayed models are listed in
Table \ref{tab:parameters} at the reference scale $\mu^2=4$~GeV$^2$.
This scale is employed in lattice calculations because it naturally
arises by the matching of the bare (lattice) operators at
$\mu_{0}^{2} =1/a^2$ ($a$ being the lattice spacing) to those in the
$\overline{\rm MS}$ scheme in continuum QCD.
It is also used in various works based on the DSE approach.
In this work, we obtained our own results at the initial scale
$\mu^2 \gtrsim 1$~GeV$^2$ and evolved them to this higher scale using
ERBL evolution at the next-to-leading order level.
The conformal coefficients $a_2$ and $a_4$,
and the moments $\langle \xi^2 \rangle$ and
$\langle \xi^4 \rangle$ along with the inverse moment
$\langle x^{-1} \rangle$ in Table \ref{tab:parameters} have for
each DA their own original values at $\mu^2=4$~GeV$^2$.
In cases where the original scale was lower, we have evolved these
quantities to the scale $\mu^2=4$~GeV$^2$ using two-loop ERBL
evolution.

We have included in the table the AdS/QCD model of the pion DA
derived in \cite{BT07} within holographic QCD.
This model reads
$
 \varphi_\pi(x, \mu^2)=(4/\sqrt{3}\pi)\sqrt{x\bar{x}}
$
and approaches at $Q^2\to \infty$ the asymptotic DA
$\varphi_{\pi}^{\rm asy}(x)=6x\bar{x}$,
while it has a very different $x$ behavior at finite
$Q^2$ \cite{BT07,BCT11}.
The conformal coefficients of this DA have been computed in
the arXiv version of \cite{Stefanis:2008zi} at the
initial scale $\mu^2\approx 1$~GeV$^2$ to obtain
$\langle \xi^2\rangle^{\rm AdS/QCD}=0.250$ and
$\langle \xi^4\rangle^{\rm AdS/QCD}=0.125$
from which the first two conformal coefficients
$a_{2}^{\rm AdS/QCD}=7/48$
and
$a_{4}^{\rm AdS/QCD}=11/192$ were determined.
Using NLO scaling relations, these coefficients have been evolved to
the reference scale $\mu^2=4$~GeV$^2$ and are given in Table
\ref{tab:parameters}.
The value of the second moment was later computed in \cite{BCT11} for
$\mu_0=1$~GeV and $\Lambda_{\rm QCD}=0.225$~GeV using LO evolution
and found to be almost the same, notably,
$\langle \xi^2 \rangle^{\rm AdS/QCD}_{\mu^2=4~{\rm GeV}^2}=0.24$.
Note that a broad pion DA $\sim \sqrt{x\bar{x}}$
was considered before in \cite{MR86,MR90} using QCD sum rules with
NLCs.
An even broader DA was discussed earlier in \cite{Dittes:1981aw},
giving
$\varphi_\pi(x) \sim (x\bar{x})^{[0.1 - 0.2]}$.
To complete this discussion, we mention that the
coefficient $a_2$ (and other parameters) for the pion DA proposed by
Chernyak and Zhitnitsky \cite{Chernyak:1983ej} has been included in
Table \ref{tab:parameters} using NLO evolution to the scale
$\mu^2=4$~GeV$^2$, but we have not displayed it in Fig.\ \ref{fig:rhoL}
because its value is outside the range of the graphs.

The most important observations from Fig.\ \ref{fig:rhoL} are the
following:
(i) There is no overlap of the $(a_2,a_4)$, or, equivalently,
$(\alpha_{-}, a_{2}^{\alpha})$, regions for the pion and
the $\rho_{\parallel}$ meson computed with NLC sum rules in
\cite{BMS01} and \cite{Pimikov:2013usa}.
(ii) While the platykurtic regime for the $\rho_{\parallel}$ DA is
entirely enclosed by the region determined with
$\lambda_{q}^2=0.4$~GeV$^2$,
the platykurtic regime for the pion DA appears as an exclave.
This is, because in order to obtain a platykurtic pion DA one has to
use the slightly larger value
$\lambda_{q}^2=0.45$~GeV$^2$.
(iii) The DSE-based DAs for the pion and the $\rho_{\parallel}$ meson
are in both panels far away from our NLC estimates.
But keep in mind that the locations of the DSE DAs in the $(a_2,a_4)$
plane are only indicative because, as we have already mentioned, their
parametrization by means of Eq.\ (\ref{eq:truncated}) is a very crude
approximation.
(iv) We observe in the $(a_2,a_4)$ plane an intriguing alignment
of the unimodal DSE-based DAs ($\pi$ and $\rho_\parallel$) along an
upward pointing ``diagonal'' and another branch of bimodal NLC-based
DAs steered downwards along $a_2 + a_4 \approx {\rm const}$.
This second ``orbit'' of DAs roughly follows the values of
$\langle x^{-1} \rangle_{\pi}/3 - 1$ evolved to the scale 4~GeV$^2$
\cite{BMS01}.
Crucially, there is a small region where both branches overlap
allowing the combination of endpoint suppression with unimodality ---
the chimera regime of our interest.
(v) On the other hand, all DAs, blended together according
to Eq.\ (\ref{eq:hybrid}), lie on the straight line with
$a_2 \cong 0.15$ and stray away from the NLC $(a_2,a_4)$ region with
decreasing values of the mixing parameter $a$.
Ultimately, i.e., for $Q^2\to \infty$, all DAs with the correct
$x$ asymptotics will evolve either along the upper ``diagonal''
(if they are unimodal) or along the lower one (if they are bimodal)
toward the asymptotic DA (\ding{117}) with $\gamma_0=0$,
as predicted by perturbative QCD.
Issues (iv) and (v) will be addressed later in full detail.

\section{Meson DA\lowercase{s} as patterns of synchronization}
\label{sec:sync}
The meson DA at a fixed scale $\mu^2$ is a distribution of $x$ values
in the interval $[0,1]$.
While a unimodal DA profile may seem more ``natural'' in appearance for
the ground state of the pion, the interpretation of a bimodal structure
causes discomfort and gives rise to debates.
So there is a desire for an explanation and rationalization of this
issue.

Recently, it was argued by one of us \cite{Stefanis:2014nla} that
in order to better understand the patterns of the pion and other
meson DAs it is useful to develop some ideas which are drawn from
the subject of synchronization of nonlinear
oscillators in the theory of complex systems --- natural and
engineered (see \cite{Strogatz2000,Acebron:2005zz} for reviews).
The nub of the idea, as Stefanis put it in
Ref.\ \cite{Stefanis:2014nla}, is to represent the $x$ values,
accessible to the meson DAs in the interval $[0,1]$, in terms of the
phases of a large number ($N\to\infty$) of interacting oscillators.
The dynamics of this kind of systems is describable in terms of the
Kuramoto model \cite{Kuramoto:sync1984} and its descendants, albeit
its specifics is not relevant for the present analysis.
What is more important is that the synchronization of the oscillator
phases, alias the longitudinal momentum fractions carried by the
valence quark vs. that of the antiquark, gives rise to the formation
of particular patterns of the $x$ distribution.
These patterns emerge from the ``organization'' of the phase spectrum
(i.e, the $x$ distribution) and reflect the specific approach used to
describe the partonic interactions in the pion bound state described
by the DA.
In other words, each particular DA profile is latent in the
underlying theoretical method and pertains to a patterned arrangement
of synchronized coupled oscillators in the Kuramoto context.

These methods can be QCD sum rules with nonlocal condensates,
as employed in this work and in \cite{BMS01,Pimikov:2013usa}
(see also \cite{BMPS11,Bakulev:2012nh}),
local condensates \cite{Chernyak:1983ej},
lightcone sum rules (LCSR)s \cite{ABOP10,ABOP12},
instanton models \cite{Dorokhov:2006xw},
approaches based on Dyson--Schwinger-equations
\cite{Chang:2013pq,Gao:2014bca} (reviewed in \cite{Cloet:2013jya}),
light-front models, e.g., \cite{Choi:2007yu,Choi:2014ifm},
holographic QCD \cite{BT07,BCT11} --- see \cite{Brodsky:2013dca} for
a recent review, etc.
Examples of $\pi$ and $\rho_\parallel$ DA profiles are depicted in
Fig.\ \ref{fig:pi-rho-DAs}.
Thus, the Sync concept provides a universal canvas to study the
characteristics of very different meson DAs without taking recourse
to a specific calculational scheme.
In particular, it puts a theoretical basis beneath the interpretation
of the bimodality of meson DAs, as we will show next.

Indeed, it was pointed out in \cite{Stefanis:2014nla} that at scales
of the order of $\mu \sim 1-2$~GeV, nonlocal condensates, which are
used to parameterize the vacuum nonlocality in terms of a nonvanishing
quark virtuality $\lambda_{q}^{2}$
cause the distribution over $x$ to flock into two clusters, giving
rise --- within a broad range of uncertainties in the midregion of $x$
--- to bimodal DAs for the pion \cite{BMS01} and the $\rho_{\parallel}$
meson \cite{Pimikov:2013usa}.
The corresponding families of DAs are shown in the form of the larger
shaded bands in Fig.\ \ref{fig:pi-rho-DAs}.
The upper panel refers to the $\pi$ and the lower one to the
$\rho_\parallel$ meson, both at the scale $\mu^2=4$~GeV$^2$
after NLO evolution from the initial scale
$\mu_{0}^{2} \sim 1$~GeV$^2$.

The bimodality strength of the BMS-type of DAs is controlled by the
nonlocality parameter $\lambda_{q}^{2}$, which endows vacuum
fluctuations with a characteristic correlation length
$\sim 1/\lambda_q$.
Lower values of $\lambda_{q}^{2}$ tend to increase the bimodality
character of the DA and reduce the value of
$\varphi_{\pi}(x=1/2)$, while larger values enhance the midregion
of $x$ driving $\varphi_{\pi}(x)$ closer to a unimodal
distributional shape.
This behavior is deeply rooted in the combined effect of the
perturbative part and the power-behaved terms in the QCD sum rule for
the moments $\langle \xi^{N} \rangle$ considered in \cite{BMS01} and
in \cite{Pimikov:2013usa}.
For $\lambda_{q}^{2}=0$, one recovers the QCD sum rules of
Chernyak-Zhitnitsky in Ref.\ \cite{Chernyak:1983ej} with an infinite
correlation length of the vacuum fluctuations.
The numerically most important term is the scalar-condensate
contribution \cite{MR86}, encountered in
(\ref{eq:SRrhoLterms}),\footnote{The expression for the scalar
quark-condensate contribution in Eq.\ (\ref{Phi-Scalar}) pertains to a
Gaussian model for the quark condensate.}
\begin{eqnarray}
  \Phi_S\left(x;M^2;\Delta\right)
&& \!\!\!\!\!\! = \! \frac{A_S}{M^4}\frac{18}{\bar\Delta\Delta^2}
       \left\{
       \theta\left(\bar x>\Delta>x\right)
       \bar x\left[x \right. \right.
\nonumber \\
&& \!+\! \left. \left.
       (\Delta-x)
       \ln\left(\bar x\right)\right]
       \! + \!  \left(\bar x\rightarrow x\right)
       \! + \! \theta(1>\Delta) \right.
\nonumber \\
&& \!\!\times \left.
         \theta\left(\Delta>x>\bar\Delta\right)
         \! \left[\bar\Delta
              +\left(\Delta-2\bar xx\right) \right. \right.
\nonumber \\
&& \!\!\times \! \left. \left.
         \ln(\Delta)\right]
         \right\} \, ,
\label{Phi-Scalar}
\end{eqnarray}
where
$
 A_S
=
 \left(8\pi\alpha_s\right/81)
 \langle \bar{q}q \rangle^2
$,
with the four-quark contribution being given by
$
 \alpha_s \langle \bar{q}q \rangle^2
=
 1.83\times 10^{-4}
$~GeV$^6$
and
$
 \langle
        \alpha_s GG
 \rangle/12\pi
=
 0.0012
$~GeV$^4$ \cite{SVZ1}.
Here, $\Delta=\lambda_{q}^{2}/2M^2$, $\bar{\Delta}\equiv 1-\Delta$,
and $M^2 \approx 1$~GeV$^2$ is the Borel parameter
$M^2\in[M_{\rm min}^{2}, M_{\rm max}^{2}]$.
The sum rule should not be sensitive to the choice of this parameter.
The procedure for minimizing the dependence on $M^2$ has been
described in \cite{MPS10} and is applied here.

Larger values of $\Delta$ shift the balance in the sum rule in favor
of the perturbative contribution which has a single mode at $x=1/2$,
thus entailing a reduction of the two peaks at $x_0=\Delta$ and
$x_0=\bar\Delta$ until they ultimately collapse into a single more
rounded peak at the center.
But despite this shift towards the central region of $x$, the tails of
the BMS DAs remain suppressed within only a small range of
theoretical uncertainties, as one also observes from
Fig.\ \ref{fig:pi-rho-DAs}.
Moreover, it was shown in \cite{MPS10} that the endpoint
behavior of the pion DA can be related to the ``decay rate'' of the
correlation length of the scalar condensate.
It is worth noting in this context that in the local version of the
condensate model, i.e., for $\lambda_{q}^{2}=0$, all nonperturbative
contributions are concentrated just at the endpoints because
$
 \Phi_{4\rm Q}^{\rm local}(x)
=
 9\left[\delta(x)+\delta(\bar{x})\right]/(M^2)^2
$.

Viewed from the Kuramoto prism, the two peaks of the BMS DAs
correspond to two distinct groups of rather strongly synchronized
oscillators with characteristic ``frequencies'' located in the
lower and upper quartiles of the $x$ distribution, respectively.
On the other hand, the tails correspond to tiny cohorts of oscillators
with natural ``frequencies'' close to the rare values $x=0$ and $x=1$
with almost nil phase-locking, while a partly synchronized arrangement
of oscillators with values around $x=1/2$ connects the two clusters
across the dip in the central region.

Consider now the implementation of DCSB to meson DAs.
It was stated in \cite{Stefanis:2014nla} that the DCSB and the
concomitant mass generation of quarks and gluons within the DSE-based
framework \cite{Chang:2013pq,Gao:2014bca} tend to enhance, both the
central region of $x$ values but also the tails of the meson DAs down
to the kinematic endpoints $x=0,1$, leading to a homogenization of the
$x$ values of the valence $\bar{q}q$ pair and to broad unimodal DAs for
all considered mesons
\cite{Chang:2013pq,Gao:2014bca,Shi:2014uwa,Shi:2015esa}.
These DAs have downward concave profiles in the whole interval
$x\in[0,1]$.
There are two variants of pion DAs derived from the DSE-based approach
\cite{Chang:2013pq}.
They were computed via a large number of moments
$\langle (x-\bar{x})^N \rangle$
$(N=50)$ and were then expressed by means of Eq.\ (\ref{eq:DSE-DA})
using two different DSE truncations at the scale $\mu^2=4$~GeV$^2$.
The associated values of the parameters
$\left(N_{\alpha}, \alpha_{-}, a_{2}^{\alpha}\right)$
and other relevant metrics
are provided in Table \ref{tab:parameters}.

One $\pi$ DA, dubbed DSE-RL, was obtained using the rainbow-ladder (RL)
approximation of the Bethe--Salpeter kernel in the DSEs, while the
other, termed DSE-DB, was derived with a DCSB-improved kernel
(abbreviated by DB), which includes nonperturbative DCSB-generated
effects that were not taken into account in the RL truncated version.
Both DAs are much broader relative to $\varphi_{\pi}^{\rm asy}$, with
the DSE-RL DA being flatter and broader than the DSE-DB DA.
The profiles of these DAs are given in the upper panel of
Fig.\ \ref{fig:pi-rho-DAs}: DSE-RL --- dashed-dotted line; DSE-DB
--- dashed line.
Also the $\rho_\parallel$ DA obtained with the DSE computational
scheme \cite{Gao:2014bca}, is a relatively broad everywhere
downward concave curve
(though less pronounced than both pion DSE DAs), as
one can see from the lower panel of Fig. \ref{fig:pi-rho-DAs}
(lower dashed line).
The broadening of the DSE DAs is a direct consequence of the
nonperturbative DCSB interactions which give rise to the dressed
quark's selfenergy --- see \cite{Cloet:2013jya} for a detailed review
of the method and further explanations.
Also note that the DSE DAs cross the NLC-based ones twice on each side
of the mean (Fig.\ \ref{fig:pi-rho-DAs}).
Thus, these DAs show a pattern of higher-lower-higher on each side,
related to their heavier tails.
This behavior can be quantified by employing the kurtosis statistic,
defined by
\begin{equation}
  \beta_{2}[\varphi]
=
  \frac{E(x-\mu[\varphi])^4}{(E(x-\mu[\varphi])^2)^2}
=
  \frac{\mu_4[\varphi]}{\sigma^4[\varphi]}
=
  \frac{\langle \xi^4 \rangle_\pi}{\left(\langle \xi^2 \rangle_\pi\right)^2} \, ,
\label{eq:kurtosis}
\end{equation}
where
\begin{equation}
  \sigma^2[\varphi]
=
  \int_{0}^{1} dx (x-\mu[\varphi])^2 \varphi(x)
=
  \frac{1}{4}\langle \xi^2\rangle_\pi
\label{eq:variance}
\end{equation}
is the variance of the distribution $\varphi(x)$.
Together with the skewness
(vanishing here but being relevant for mesons composed of light
and heavy quarks like the kaon)
\begin{equation}
  \gamma_{1}[\varphi]
=
  \frac{E(x-\mu[\varphi])^3}{\sigma^3[\varphi]} \, ,
\label{eq:skewness}
\end{equation}
it describes  the central tendency, variability, and shape of a
distribution.
In particular the kurtosis serves to measure the peakedness in the
central region of a distribution against the flatness of its tails.
As one sees from Table \ref{tab:parameters}, the unimodal,
downward concave DSE-based DAs
can be ordered as follows:
$
  \beta_{2}^{\rho_{\parallel}}
>
  \beta_{2}^{\pi_{\rm DB}}
>
  \beta_{2}^{\pi_{\rm RL}}
$.
This result confirms a similar qualitative statement in
\cite{Gao:2014bca}.

Such broad DA morphologies describe a loosely synchronized
assortment of oscillators spread over the entire range of their
native ``frequencies'' in $x\in [0,1]$.
The enhancement of the generic midregion, which corresponds
--- from a physical perspective --- to ``egalitarian'' partonic
configurations in the pion in which the valence quark and the
valence antiquark carry comparable fractions of longitudinal
momentum, may be welcome.
But at the same time DSE DAs also overestimate the weight of
``aristocratic'' configurations with low probability occurrence
in which a single valence parton takes the lion's share of the
momentum with $x\to 1$ or $\bar{x}\to 1$ to go far-off shell.
Because these are more specific and rare configurations of the
dispersion of the valence parton's longitudinal momentum far away
from the typical values around the mean $\mu=1/2$, tail enhancement
is at odds with our understanding of the QCD description of
the pion bound state based on off-shell gluon exchanges
(see \cite{SSK99,SSK00} and \cite{BCT11} for explanations)
and, as we have seen, leads to more variation in the oscillator
``frequencies'', thus entailing less synchronization.
Note that the broader and flatter a unimodal downward
concave $x$ distribution is, the closer to random the oscillator
phases will be.

Comparison in earlier works, e.g.,
\cite{BMS05lat,Bakulev:2012nh,Stefanis:2014yha},
of predictions for the pion-photon transition form factor obtained
within the LCSR framework with all existing data, indicates that strict
QCD scaling behavior at high $Q^2$ is very sensitive to the
endpoint-behavior of the pion DA.
This behavior is intimately related to the inverse moment
$\langle x^{-1}\rangle_{\pi}$.
Recalling Eq.\ (\ref{eq:diagonal}), this implies that the conformal
coefficients have to balance each other in such a way so that
the excess over $\langle x^{-1}\rangle_{\pi}^{\rm asy}=3$ is not
too large.
Otherwise, the form factor would overshoot the data.
And in fact, using the DSE DAs as nonperturbative input in a
LCSR-framework, we have shown \cite{Stefanis:2014yha} that the
predictions obtained herewith exceed the asymptotic limit
even at the highest momenta around 40~GeV$^2$ probed in
current experiments \cite{BaBar09,Belle12}.
This is also obvious from Table \ref{tab:parameters}.
The inverse moment
\begin{equation}
  \langle x^{-1} \rangle_{\pi}
=
  \frac{1+2\alpha_{-}}{\alpha_{-}}\left(
                                        1 + a_{2}^{\alpha} + \ldots
                                  \right) \, ,
\label{eq:inv-mom--alpha}
\end{equation}
obtained with both pion DSE DAs, has very large values to be
compatible with the data.
We mention in this context that a recent analysis
\cite{Swarnkar:2015osa} of the meson structure in lightfront
holographic QCD, which employs a broad unimodal $\pi$ DA, finds
predictions for the pion-photon transition form factor
that are in quite poor agreement with all experimental data
(see Fig. 18 there).
But notice that a subsequent DSE-based computation
\cite{Raya:2015gva} of the TFF, in which the role of the inverse moment is
not so crucial, finds good agreement with the CELLO, CLEO, and Belle data
in the entire domain of spacelike momenta.

\begin{figure}[t]
\centering
\includegraphics[width=0.45\textwidth]{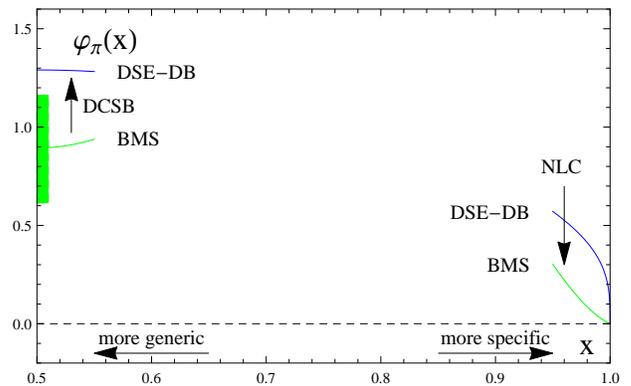} 
\caption{(color online). Illustrating the key antithetic effects in the
pion distribution amplitude in the interval $[0.5,1]$ (mirror
graph in the interval $[0,0.5]$ not shown).
The lower (green) line denotes the BMS DA, with the range of such DAs
being indicated by the green vertical strip (cf. green shaded band in
the upper panel of Fig.\ \ref{fig:pi-rho-DAs}).
The upper (blue) line marks the DSE-DB DA in the central $x$ region.
The unimodal DSE-DB DA shows enhancement of the tails and the
midregion around $x=0.5$, while the BMS $\pi$ DAs are characterized
by suppression of the endpoint region $x=1$.
The corresponding main trends are indicated by vertical arrows:
DCSB --- enhancement;
NLC --- suppression.
\label{fig:oxymoron}}
\end{figure}

In the Sync analogy, the NLC within our approach causes a generic
clustering of the DA into two clusters liaised with massive endpoint
suppression --- within the range of intrinsic theoretical uncertainties ---
and only a moderate reduction of the DA in the central region, which is
controlled by the strength of the nonlocality parameter as discussed
above in connection with Eq.\ (\ref{Phi-Scalar}).
Thus, configurations with a highly asymmetric dispersion of the
valence parton's longitudinal momentum are suppressed.
It is highly unlikely that the pion can remain intact and rebound
as a whole for such partonic configurations.
In the Sync analogy they would play the role of very idiosyncratic
oscillators incapable to synchronize, being either too slow (with
``frequency'' values close to 0) or too fast with ``frequencies''
tending to unity.
In this sense, the NLC acts like a negative feedback opposing the
excessive DCSB-induced enhancement in the endpoint regions by
turning off the corresponding oscillators.
A figurative explanation of these issues is provided in
Fig.\ \ref{fig:oxymoron} in which the two main antithetic effects in
the $x$ behavior of the valence quark in the pion DA are illustrated.
This is how Stefanis envisioned and mapped out the Sync properties
of the BMS-like and DSE-like pion DAs in \cite{Stefanis:2014nla}.
So there are, it seems, two very distinct DA patterns with telltale
signatures that include, but are not limited to, the pion case:
One is related to NLCs, which encode particle correlations in the range
$1/\lambda_q \sim 0.3$~fermi, while the other implements DCSB which
causes the dynamical generation of quark masses and entails dressing
of the quark propagator describing the confined quark in the pion.
Both effects are manifestations of confinement and neither exists
in isolation --- see \cite{Brambilla:2014jmp} for a recent review of
strong-interaction dynamics.
Unfortunately, they cannot be described simultaneously within a single
analytic approach at present.
Thus, for the time being, there are two physical
paradigms with their own computational techniques, each applying only
within its own sphere of acceptance and validity.

\section{Synthetic meson DA\lowercase{s}}
\label{sec:synthetic-DA}
There are basically two options:
(i) Either the $x$ distribution of the pion DA is described by a single
DA over the whole range of values in the interval $x\in [0,1]$, or (ii)
the ``true'' pion DA is rather a mixture of two different DAs, one
better applicable to the central region and the other controlling the
tails.
The first option was discussed above and at length in the literature.
Following the second scenario, Stefanis proposed in
\cite{Stefanis:2014nla} a synthetic DA of the form
\begin{equation}
  \varphi_{\pi}^\text{true}(x, a)
=
 a\varphi_{\pi}^\text{BMS}(x) + (1-a)\varphi_{\pi}^\text{DSE}(x) \ ,
\label{eq:hybrid}
\end{equation}
where $a$ is a mixing parameter with values within the interval
$[0,1]$.
Mixtures of the form of Eq.\ (\ref{eq:hybrid}) are quite common in
statistics when a single distribution, like the Gaussian distribution
function, the Poisson distribution, etc.,
has to be combined with a distribution with a different type of
mathematical behavior in the tails, e.g., the generalized Pareto
distribution.
In fact, hybrid-like DAs have been constructed and profoundly studied
by Bergmann and Stefanis long ago for the nucleon
\cite{Stefanis:1992nw,Bergmann:1993eu,Bergmann:1993rz}
and also for the $\Delta^{+}(1232)$ resonance \cite{Stefanis:1992pi}
(termed ``heterotic'' DAs), although they were motivated by other
concerns.
A comprehensive and detailed review of such baryon DAs has been given
in \cite{Stefanis:1999wy}.
In the present case, the coexistence of distinct domains of
oscillators, some coherent and phase-locked (BMS peaks), and others
which describe unsynchronized oscillators (heavy tails of the DSE
DA), would give rise to a so-called chimera state
\cite{Kuramoto:2002,Abrams:2004}.

It was argued in \cite{Stefanis:2014nla} that for values of the
mixing coefficient $a$ close to 1, the synthetic DA would still
belong to the family of pion BMS DAs shown in terms of the wide shaded
band in Fig.\ \ref{fig:pi-rho-DAs}.
More generally, the synthesized DA is supposed to encapsulate both
manifestations of QCD confinement, reflecting the perpetual balance
which arises from the appropriate combination of the two basic
effects, one associated with NLC formation, via QCD
sum rules (BMS-DA \cite{BMS01}), the other being the result of DCSB,
expressed in terms of a DA computed within the DSE-based approach
\cite{Chang:2013pq}.
A proper combination has to balance the enhancement impact of DCSB
against the suppression due to NLC, as exposed in
Fig.\ \ref{fig:oxymoron}.

Under ERBL evolution the synthetic DA would develop at $Q^2\to \infty$
to the asymptotic DA which represents the most synchronized
configuration of the valence $\bar{q}q$ pair within the
pion being still a bound state \cite{Stefanis:2014nla} after all
quark-gluon interactions have died out.
This is also evident from Table \ref{tab:parameters} from which we see
that the asymptotic DA has the largest kurtosis, i.e.,
$\varphi_{\rm asy}$ is the most leptokurtic meson DA.
In this paper, we work out Eq.\ (\ref{eq:hybrid}) in more certain
terms and exploit the whole range of possible values of $a$.
From the synchronization point of view, a synthetic DA represents an
attempt of combining ensembles of synchronized and unsynchronized
(but otherwise identical) oscillators in order to enhance or suppress
particular frequency values amounting to a chimera state.
The question is whether the simple one-parametric design of
Eq.\ (\ref{eq:hybrid}) is indeed capable of providing the desired
properties addressed above in the adjunct discussion of
Fig.\ \ref{fig:oxymoron}.

\begin{figure}[t]
\centering
\includegraphics[width=0.45\textwidth]{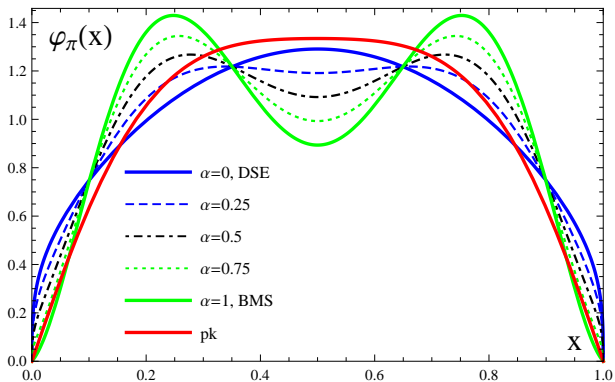} 
\caption{(color online). Synthetic pion DAs
obtained with
Eq.\ (\ref{eq:hybrid}) at the scale $\mu^2=4$~GeV$^2$ for various
values of the mixing parameter $a\in [0,1]$.
The shorttailed platykurtic (pk) DA is shown by the upper red solid
line for comparison; it does not belong to this class of DAs
(see text).
\label{fig:DA-mixture}}
\end{figure}

Figure \ref{fig:DA-mixture} shows various samples of synthesized DAs
obtained with different values of the mixing parameter between 0 and 1.
As one clearly sees from these plots, when $a$ is larger than, say,
0.75, the main characteristics of the BMS DAs persist.
This is also obvious from the location of this DA,
denoted by the symbol (\ding{174}), in the plane $(a_2,a_4)$ in the
left panel of Fig.\ \ref{fig:rhoL}.
However, for small $a$ values close to 0.25 and below, the obtained
DA profiles show no endpoint suppression, as desired, but exhibit
instead tail enhancement, inherited to them by the DSE DA, while the
profile is still bimodal, see \ding{172} ($a=0.25$) and
\ding{173} ($a=0.5$) in the same figure.
This means that the bimodality of the DA prevails from large down to
quite small values of the mixing parameter while at the same time
for these small values of $a$ the tails of the DA get strongly
enhanced.
Hence, Eq.\ (\ref{eq:hybrid}) cannot supply a pion DA which combines
unimodality in the central region with suppression of the tails,
despite the fact that it generates DAs with the same value of $a_2$,
see, Fig.\ \ref{fig:rhoL}, left panel.
Nor can this be realized via ERBL evolution.\footnote{Strictly
speaking, evolution in NLO gives a logarithmic modification
$\sim (\alpha_s/(4\pi))C_{\rm F}\ln^2[(1-x)/x]$ \cite{Mul95} which
obviously affects the endpoint behavior of the meson DA ---
independently of its shape --- albeit endpoint-enhanced DAs
receive larger NLO corrections \cite{Mul94}.
However, given that at scales $\mu\geq 1$~GeV$^2$ the running coupling
is already sufficiently small, this effect can be safely neglected,
see \cite{BMS01}.}
In order to embody endpoint suppression of the pion DA, one has to
build it in right from the start, resorting to QCD sum rules with NLCs
and looking for DAs which would mimic the characteristics of the
DSE DAs in the central region, while preserving suppression of the
tails in compliance with Fig.\ \ref{fig:oxymoron}.
To achieve this goal we have to keep in mind that if mass is moved
from the shoulders to the center of a distribution, then one
has to compensate the accompanying movement of mass to the tails,
leaving the variance almost unchanged but increasing the kurtosis.

The existence of such a chimera DA for the pion, which binds these
diverse aspects of coherence and incoherence into a single DA, was
first discussed in \cite{Stefanis:2014nla} and predictions for the
pion-photon transition form factor
$Q^2F^{\gamma^*\gamma\pi^{0}}(Q^2)$
were presented which are fully compliant with all experimental data
compatible with QCD scaling.
The overall quality of these predictions resembles that of the
BMS-type DAs \cite{Stefanis:2014nla,Stefanis:2014yha}.
This can be traced back to the fact that they both lead to an inverse
moment with just the appropriate size in order to agree with the
$Q^2F^{\gamma^{*}\gamma\pi^{0}}(Q^2)$ data.
In fact, a brand-new simultaneous fit to the
CLEO \cite{CLEO98} and Belle \cite{Belle12} data in \cite{Zhong:2015nxa}
favors a profile of the pion DA which is very close to the platykurtic
one (see Fig.\ \ref{fig:pi-rho-DAs} and Table \ref{tab:parameters}).
On the other hand, DAs with downward concave shapes in the
entire interval $x\in[0,1]$, will tend to overestimate most data of
the Belle Collaboration \cite{Belle12}, the reason being that they
give rise to large inverse-moment values (Table \ref{tab:parameters})
and \cite{Stefanis:2014nla,Stefanis:2014yha}
(see also \cite{Swarnkar:2015osa}).
[The opposite behavior was found in \cite{Raya:2015gva}, as already
mentioned.]

In this work, we determine a whole domain of such chimera $\pi$ DAs
employing our NLC technology and allowing for a slightly larger value
of the quark virtuality, viz., $\lambda_{q}^{2}=0.45$~GeV$^2$.
The core area of these DAs is illustrated in the upper panel of
Fig.\ \ref{fig:pi-rho-DAs} in the form of a narrow strip in red color.
The corresponding parameters and the range of theoretical uncertainties
for both used Gegenbauer parametrizations are given at the reference
scale $\mu^2=4$~GeV$^2$ in Table \ref{tab:parameters}.
This table also includes the brand-new lattice estimates for the second
moment $\langle \xi^2 \rangle$ from \cite{Braun:2015axa} at the same
scale using the $\overline{\rm MS}$ scheme.

Note that this $a_2$ value was not calculated from the second moment
$ \langle \xi^2 \rangle$ via Eq.\ (\ref{eq:conf-coeff-a2}) at finite
lattice spacing, taking subsequently the continuum limit.
Instead, the value of $a_2$ was calculated directly on the lattice and
was then extrapolated to the continuum limit via
$
 \langle \xi^2 \rangle_{a\neq 0}^{\overline{\rm MS}}
\Rightarrow
 a_{2}^{\overline{\rm MS}}|_{a\neq 0}
\Rightarrow
 a_{2}^{\overline{\rm MS}}|_{a=0}
$.
This implies that Eq.\ (\ref{eq:conf-coeff-a2}) is broken by lattice
artifacts.
A small variation in the lattice spacing around $6\%$ may result in
an increase of $a_2$ of the order of $25-30\%$ \cite{Braun:2015axa}.
The final result at $\mu^2=4$~GeV$^2$ reads
$a_{2}^{\overline{\rm MS}}|_{a=0}=0.1364(154)(145)$,
while the reported value of the second moment is
$\langle \xi^2 \rangle_{\pi}^{\rm lat}=0.2361(41)(39)$.
The first error is statistical and originates from the chiral
expansion, whereas the second one pertains to the uncertainties of the
renormalization factors.
It agrees within errors with
$\langle \xi^2 \rangle_{\pi}^{\rm BMS}=0.248\{_{0.233}^{0.264}$
\cite{Stefanis:2008zi}
and also with
$\langle \xi^2 \rangle_{\pi}^{\rm pk}=0.220\{_{0.213}^{0.229}$
(see Table \ref{tab:parameters}).
In contrast, while the lattice
estimate $a_{2}^{\rm lat}$ agrees with the BMS coefficient
$a_{2}^{\rm BMS}$, determined in the year 2001 \cite{BMS01}
(see Table \ref{tab:parameters}),
it turns out to be larger than $a_{2}^{\rm pk}$.
But one should be cautious.
Extracting $a_2$ via
$
 \langle \xi^2 \rangle_{a\neq 0}^{\overline{\rm MS}}
\Rightarrow
\langle \xi^2 \rangle_{a= 0}^{\overline{\rm MS}}
\Rightarrow
 a_{2}^{\overline{\rm MS}}|_{a=0}
$, one would obtain $a_{2}^{\rm lat}=0.105$, which is indeed compatible
with the range of the platykurtic $a_2$ values.
Thus, one cannot exclude the influence of significant discretization
effects that would require simulations at smaller lattice spacings
of the order of $a \sim 0.04$~fm \cite{Braun:2015axa}.

Be that as it may, one should recall that the second moment
$\langle \xi^2 \rangle_{\pi}$ is related to the
variance of the DA given by Eq.\ (\ref{eq:variance}).
This statistic is not sufficient to draw any conclusions about
the shape of the distribution in the central region.
Indeed, as one observes from Table \ref{tab:parameters}, the unimodal
DSE-DB pion DA yields a conformal coefficient
$a_{2}^{\rm DSE-DB}=0.149$ which fully agrees with the new lattice
result but also with $a_2^{\rm BMS}$.
We note that this is valid for the second moment as well, which has
the value
$\langle \xi^2 \rangle_{\pi}^{\rm DSE-DB}=0.250$, (cf.
(\ref{eq:conf-coeff-a2})) and thus almost coincides with the second
moment of the pion BMS DA given above, being also close to
$\langle \xi^2 \rangle_{\pi}^{\rm pk}$.
On the other hand, the fourth moment $\langle \xi^4 \rangle$ and the
conformal coefficient $a_4$ of the DSE and the BMS (pk) DAs are
different in value and sign, respectively ---
see Table \ref{tab:parameters}.
What is far more significant is the fact that, as it is evident from
the left panel of Fig.\ \ref{fig:rhoL}, \emph{all} DAs lying on
the straight vertical line at $a_2\cong 0.15$ agree equally well
with the new lattice estimate for $a_2$.
This makes it apparent that a single lattice constraint cannot fix the
profile of the pion DA uniquely, however precise it may be.

The chimera DAs have shorttailed platykurtic profiles and
overlap with the DSE-DB DA in the midregion of $x$ but descend at the
endpoints at low angle to zero, similar to a typical BMS DA.
As one observes from Fig.\ \ref{fig:rhoL} (both panels), there is an
imbrication of the platykurtic regimes (small rectangles in
light-green color) with the domains of the bimodal pion DAs obtained
with NLC sum rules for the quark virtuality
$\lambda_{q}^2=0.45$~GeV$^2$ (transparent rectangles bounded by a
dashed line).
For this value the conformal coefficients for
$\varphi_{\pi}^{\rm BMS}(x)$ at $\mu^2=4$~GeV$^2$ read
$a_2=0.12$ and $a_4=-0.06$ while the inverse moment is
$\langle x^{-1}\rangle_{\pi}^{{\rm BMS}(\lambda_{q}^{2}=0.45~{\rm GeV}^2)}=3.18$,
a value which agrees well with
$\langle x^{-1}\rangle_{\pi}^{\rm pk}=3.13$
in Table \ref{tab:parameters}.
The prediction for $Q^2F^{\gamma^{*}\gamma\pi^{0}}(Q^2)$ obtained with
the shorttailed platykurtic $\pi$ DA appears in line with all data of
the Belle \cite{Belle12} and the BaBar Collaboration \cite{BaBar09}
compatible with QCD scaling \cite{Stefanis:2014nla,Stefanis:2014yha}.
These unique features of the pk pion DA look indeed very attractive.
But is it more than mere coincidence or can it provide a general
mode of accessing meson DAs and offer a deeper perspective on meson
DAs in general?

To this end, we turn our attention to the $\rho$ meson case and
attempt to determine a platykurtic regime for the $\rho_\parallel$
DA using as a selector the behavior illustrated in
Fig.\ \ref{fig:oxymoron}.
Evaluating the sum rule in Eq.\ (\ref{eq:SRrhoLconden}), we compute the
reliability range of the conformal coefficients up to the order $N=10$
by first determining the central moments
$\langle \xi^N \rangle_{\rho_\parallel}$
of the same order.
Their values at the initial scale $\mu^2\gtrsim 1$~GeV$^2$ can be found
in \cite{Pimikov:2013usa}.
Also the corresponding values of the $\rho_{\parallel}^\prime$ meson
are given there together with the conformal coefficients.
We will not repeat these details here.
We concentrate instead on our primary goal to extract a platykurtic
domain of these parameters.
It turns out that this is possible even for the somewhat smaller value
of the quark virtuality $\lambda_{q}^2=0.4$~GeV$^2$ used originally
for the extraction of the pion DA in \cite{BMS01}.
The extracted domains are shown in Fig.\ \ref{fig:rhoL} in the form of
the red slanted rectangles surrounded by the larger blue bands of
coefficient values computed with the NLC sum rules.
The left panel displays the results for the first two conformal
coefficients $a_2$ and $a_4$, whereas the right panel provides the
areas of the coefficients $\alpha_{-}$ and $a_{2}^{\alpha}$.
In both graphics the platykurtic model for the $\rho_\parallel$ DA is
denoted by the symbol \ding{74}.
The values of all these parameters, accompanied by their intrinsic
errors, are compiled in Table \ref{tab:parameters}, while the
platykurtic $\rho_\parallel$ DA profiles are displayed in the form
of a narrow red strip in the lower panel of Fig.\ \ref{fig:pi-rho-DAs}.
For the sake of direct comparison with the DSE results, all graphics
and the values in the table are given at the reference scale
$\mu^2=4$~GeV$^2$ after two-loop evolution.
One immediately observes from this figure that, similar to the pion
case, the platykurtic $\rho_\parallel$ DA has a single rounded central
peak bearing endpoint suppression relative to the $\rho_\parallel$ DA
obtained within the DSE-based approach \cite{Gao:2014bca}.
In comparison to the platykurtic DA of the pion, it features a slightly
narrower profile with the kurtosis value
$
 \beta_{2}^{\rho_{\parallel}^{\rm pk}}
>
 \beta_{2}^{\pi^{\rm pk}}
 $.
Tangible consequences of the platykurtic $\rho_\parallel$ DA will be
studied elsewhere.

\section{Conclusions}
\label{sec:concl}
We have performed an intensive study of the pion and the
$\rho_\parallel$ DAs within QCD, fortified with the knowledge of
synchronization concepts used in the description of complex systems.
These concepts provide a unifying rationale of \textit{how} the various
DA profiles emerge instead of asking \textit{why} they should have a
particular shape, thus avoiding descriptive comparisons of DAs obtained
with unrelated theoretical frameworks.
Furthermore, guided by these concepts, we have used controlled theory
tools to obtain a new kind of chimera DAs for the pion and the
$\rho_\parallel$ meson using QCD sum rules with NLCs.
These DAs are capable of mingling in situ the best of both worlds ---
endpoint suppression via NLC and unimodality due to DCSB, giving rise
to shorttailed platykurtic profiles and realizing the scenario
illustrated in Fig.\ \ref{fig:oxymoron}, while preserving the
asymptotic $x$ behavior predicted by perturbative QCD.
In the Sync picture, they correspond to a vast number of phase-locked
oscillators between the lower and the upper quartile of the $x$
distribution, whereas oscillators with extremely high or low
``frequencies'', located close to the tails $x=0,1$, are in limbo.
While the characteristics of these new DAs in the central $x$ region
resemble the gross behavior of DSE-based DAs, their suppressed tails
are following the same trend as the BMS DAs.
In mathematical terms, the BMS-like DAs and the platykurtic ones are
very different as regards their profiles (Fig.\ \ref{fig:pi-rho-DAs})
and Gegenbauer coefficients (Fig.\ \ref{fig:rhoL} and
Table \ref{tab:parameters}).
But from the NLC point of view, the bimodal BMS $\pi$ DAs and the
bimodal $\rho_\parallel$ DA of \cite{Pimikov:2013usa}, which
in the Sync analogy correspond to two clusters of synchronized
oscillators, are on the same theoretical footing as the unimodal
shorttailed platykurtic DAs for these mesons, which unite
the phase-locked oscillators in a single group.
Moreover, in the pion case they yield coinciding predictions for the
pion-photon transition form factor which agree well with all
available experimental data compatible with QCD scaling above
$\sim 9$~GeV$^2$.
Given all these results, we don't want to stretch the importance of
unimodality too far.

Too broad DAs with downward concave profiles encompassing
the tails, as those derived for mesons with the aid of DSEs
\cite{Chang:2013pq,Gao:2014bca}, imply that there is no particular
$x$ value standing out because even the remote regions close to the
endpoints $x=0,1$ have a significant weight almost comparable
to that of the central region --- especially the DSE-RL pion DA.
The extreme case of a flat-top DA, like
$
 \varphi_{\pi}^\text{flat-top}(x)
=
 \Gamma(2(\alpha+1)) [\Gamma^{2}(\alpha+1)]^{-1} (x\bar{x})^{\alpha}
$
with $\alpha=0.1$ \cite{MPS10},
translates into a population of oscillators with a very strong
variation of native ``frequencies'' so that these oscillators can
hardly synchronize and as a result phase locking diminishes.
Physically, this kind of $x$ distribution comprises extremely
asymmetric partonic configurations that can spoil scale locality and
thus collinear factorization.
On the experimental side, flat-top DAs yield predictions for the scaled
pion-photon transition form factor which have a tendency to increase
with $Q^2$ --- at least in the domain of currently accessible
momentum values in the range 10-40~GeV$^2$ where one would expect
scaling to be visible \cite{Bakulev:2012nh}.
The high-$Q^2$ data of the BaBar Collaboration \cite{BaBar09} indicate
such a trend, but are not supported by the Belle data \cite{Belle12}
in the same region.
The next-generation experiments to measure the pion-photon transition
form factor with the Belle II detector at the upgraded KEKB accelerator
(SuperKEKB) in Japan and more precise data on the electromagnetic
pion form factor expected at the Jefferson Laboratory (JLab) after its
upgrade will provide extraordinary tools to test our predictions and
assertions.
\bigskip

\acknowledgments
We thank Sergey Mikhailov for collaboration on various aspects related
to the present work and for useful discussions and comments.
A.P. thanks Pengming Zhang for the warm hospitality and support
at the Institute of Modern Physics of the Chinese Academy of Sciences,
where the final stage of this work was carried out.
This work was partially supported by the Heisenberg--Landau
Program (Grant 2015), the Russian Foundation for Fundamental Research
under Grants No.\ 14-01-00647 and No.\ 15-52-04023, the
JINR-BelRFFR grant F14D-007,
the Major State Basic Research Development Program in China (No.
2015CB856903), and the National Natural Science Foundation of
China (Grant No. 11575254 and 11175215).


\end{document}